\documentclass[11pt,a4paper]{article}

\usepackage[utf8]{inputenc}
\usepackage[T1]{fontenc}
\usepackage{amsmath,amssymb,amsfonts}
\usepackage{graphicx}
\usepackage{booktabs}
\usepackage[colorlinks=true,linkcolor=blue,citecolor=blue,urlcolor=blue]{hyperref}
\usepackage{cleveref}
\usepackage{natbib}
\usepackage{xcolor}
\usepackage{caption}
\usepackage{subcaption}
\usepackage{float}
\usepackage[margin=2.5cm]{geometry}
\newcommand{\xtrue}{\tilde{x}}
\newcommand{\xhat}{\hat{x}}
\newcommand{\mhat}{\hat{m}}
\newcommand{\chat}{\hat{c}}

\title{The IQ--Motion Confound in Multi-Site Autism fMRI\\May Be Inflated by Site-Correlated Measurement Uncertainty}
\author{Kareem Soliman\\
  Independent Researcher, Sydney, Australia\\
  \texttt{kareem.soliman@outlook.com.au}\\
  ORCID: 0009-0006-7204-1129
}
\date{April 2026}

\begin{document}

\maketitle

\begin{abstract}
Multi-site autism neuroimaging studies routinely control for the confound between full-scale IQ and head motion by regressing framewise displacement against IQ scores and removing shared variance. This procedure assumes that ordinary least squares (OLS) provides an unbiased estimate of the confound magnitude. We tested this assumption on the ABIDE-I phenotypic dataset ($n = 935$ subjects across 19 international scanning sites) using Probability Cloud Regression, an errors-in-variables (EIV) estimator that models per-observation measurement uncertainty in both variables. IQ measurement error was derived from published Wechsler test-retest reliability coefficients; response-side uncertainty was represented by a site-level proxy equal to the within-site standard deviation of mean framewise displacement. Three findings emerged. First, OLS overestimates the IQ-motion slope by a factor of 4.67 relative to the EIV-corrected estimate when the bias factor is computed from the full-precision fitted coefficients ($\beta_{\mathrm{OLS}} = -0.00125$, $\beta_{\mathrm{EIV}} = -0.00027$ mm per IQ point after rounding for display). Second, under leave-site-out cross-validation a single pooled predictor of raw FD produces negative out-of-sample $R^2$ at all 19 sites (overall $R^2 = -0.074$), indicating that the pooled predictor does not transport cleanly across sites once site information is removed. Third, the direction of the EIV-corrected slope is robust across all 64 configurations of an $8 \times 8$ sensitivity grid spanning 12-fold ranges of each noise parameter. These results suggest that pooled OLS may overstate the IQ-motion association in ABIDE-I, but direct downstream consequences for motion-correction pipelines remain to be quantified using raw motion traces and connectivity-level re-analysis. Formal EIV methods appear to remain uncommon in multi-site neuroimaging confound estimation.
\end{abstract}


% ============================================================
\section{Introduction}
\label{sec:introduction}
% ============================================================

Multi-site neuroimaging consortia have transformed the study of autism spectrum conditions by pooling data across dozens of scanning sites to achieve sample sizes that no single centre could provide alone. The Autism Brain Imaging Data Exchange (ABIDE) released phenotypic and imaging data from over 1,100 participants across 20 international sites \citep{DiMartino2014}, enabling hundreds of subsequent analyses. But pooling introduces a problem that the field has spent a decade trying to solve: head motion during fMRI acquisition systematically biases measures of functional connectivity, and motion correlates with the clinical and demographic variables that multi-site studies exist to examine.

Three independent groups identified this problem simultaneously. \citet{Power2012} introduced framewise displacement (FD) as the standard motion metric and proposed volume censoring (``scrubbing'') to mitigate its effects. \citet{Satterthwaite2012} demonstrated that motion effects are particularly problematic in developmental populations, where motion correlates with age and clinical status. \citet{VanDijk2012} replicated the distance-dependent motion bias in an independent sample. Together, these papers established that even sub-millimetre head motion creates spurious short-range connectivity increases and long-range connectivity decreases, and that failure to account for motion produces false-positive group differences.

The field responded with progressively sophisticated correction methods. \citet{Satterthwaite2013} developed high-parameter confound regression models. \citet{Siegel2014} tested more stringent FD thresholds. \citet{Ciric2017} benchmarked 14 denoising pipelines and found that top-performing models include global signal regression. \citet{Ciric2018} published a comprehensive protocol that can reduce motion-related variance to near zero. Standardised preprocessing pipelines such as fMRIPrep \citep{Esteban2019} have further reduced analytic variability. In parallel, statistical harmonisation methods adapted from genomics, notably ComBat \citep{Johnson2007}, have been applied to remove site effects from derived neuroimaging measures \citep{Fortin2018, Yu2018}.

All of this work addresses motion through one of two channels: preprocessing (removing motion-contaminated signal before analysis) or harmonisation (modelling site effects in derived measures). But both channels share a hidden assumption: that the regression model used to \emph{estimate} the confound relationship is unbiased. Every scrubbing threshold, every confound regressor, every decision about which subjects to exclude is calibrated against an OLS estimate of how strongly motion relates to the variable of interest. If that OLS estimate is itself biased, then every downstream correction inherits the bias.

OLS assumes that predictors are measured without error. When both variables carry measurement noise, the OLS slope is biased. Classical errors-in-variables (EIV) theory predicts attenuation toward zero: the estimated slope is smaller in magnitude than the true slope \citep{Fuller1987, Carroll2006}. But this classical result assumes that the measurement error is independent of the true relationship across subgroups. When measurement precision and apparent effect size co-vary across subgroups, the pooled bias can reverse direction: OLS may \emph{overestimate} rather than attenuate the pooled slope. In the present setting, noisier sites also exhibit steeper apparent slopes, so subgroup structure can inflate the pooled estimator rather than shrink it toward zero.

The analytic variability problem identified by \citet{BotvinikNezer2020}, in which 70 teams analysing identical data reached substantially different conclusions, established that researcher degrees of freedom are a major source of irreproducibility. We identify an additional, orthogonal source of bias: the regression model itself. Even when all analytic choices are held constant, an OLS-based confound estimate is biased whenever both variables carry site-dependent measurement uncertainty.

In this paper, we apply errors-in-variables regression to the ABIDE-I phenotypic dataset to re-estimate the IQ-motion confound. We use Probability Cloud Regression (PCR), an EM-based EIV estimator that models each observation as a 2D Gaussian probability cloud with per-observation measurement uncertainties. IQ measurement error is derived from age-appropriate Wechsler test-retest reliability coefficients \citep{Wechsler2003, Wechsler2008}; response-side uncertainty is represented by a site-level proxy computed empirically per site. Our key finding is that OLS overestimates the IQ-motion slope by a factor of 4.67 under this proxy uncertainty model, and that a single pooled predictor of raw FD does not transport cleanly across sites under leave-site-out cross-validation.

% ============================================================
\section{Related Work}
\label{sec:related_work}
% ============================================================

\subsection{Motion Artifact Correction}

The motion artifact literature has focused on two complementary strategies: identifying contaminated volumes and removing their influence \citep{Power2012, Siegel2014}, and regressing out motion-related variance using nuisance covariates \citep{Satterthwaite2013, Ciric2017}. Both strategies require an estimate of how strongly motion relates to the imaging measure of interest, and both use OLS for that estimate. \citet{Ciric2018} provided a comprehensive protocol for motion denoising that can reduce motion-related variance to near zero, but this operates on the preprocessing side rather than on the regression model used downstream. Our work is complementary: we address the bias in the regression model itself, which operates even after perfect preprocessing.

\subsection{Multi-Site Harmonisation}

ComBat \citep{Johnson2007}, originally developed for batch effects in microarray data, has been adapted for neuroimaging applications including DTI \citep{Fortin2017} and cortical thickness \citep{Fortin2018}. \citet{Yu2018} applied ComBat to functional connectivity and found that site effects persist even with phantom-matched acquisition protocols. These methods harmonise derived neuroimaging measures by modelling additive and multiplicative site effects. Our approach addresses a different problem: the bias in the regression model used to estimate confound relationships between phenotypic variables when measurement error is site-correlated. ComBat and EIV correction are complementary, not competing.

\subsection{Errors-in-Variables Methods}

The statistical theory of errors-in-variables regression is well established \citep{Deming1943, Fuller1987, Carroll2006}. Classical results show that OLS is attenuated toward zero when predictors carry measurement error, and that the bias can be corrected when the error variance is known or estimable. Methods including regression calibration, SIMEX, and orthogonal distance regression (ODR) are available but rarely used outside of their home disciplines. \citet{Carroll2006} identified an adoption barrier: practitioners avoid EIV methods because they require knowledge of the predictor error variance, which is often unavailable. In neuroimaging, formal EIV regression still appears to be rarely used for confound estimation. The closest related neuroimaging literature focuses on measurement issues in region-of-interest analysis or on systematic errors in the BOLD signal rather than on confound regression itself.

% ============================================================
\section{Method}
\label{sec:method}
% ============================================================

\subsection{Data}

We used the ABIDE-I phenotypic dataset \citep{DiMartino2014}, publicly available from the Preprocessed Connectomes Project. The original dataset contains 1,112 subjects. We excluded subjects with missing full-scale IQ (FIQ), missing mean framewise displacement, missing age, FIQ $\leq 40$ (below the floor of reliable IQ measurement), or a quality control rating of ``fail'' from rater 1. The final sample comprised $n = 935$ subjects (449 with autism spectrum conditions, 486 controls) across 19 of the 20 ABIDE-I sites. Sample sizes per site ranged from 5 (SBL) to 174 (NYU). ASD and control groups were pooled because the IQ-motion confound estimation procedures examined here are typically applied to the combined sample in the published literature, and the present analysis targets the bias in that pooled estimation procedure rather than diagnostic group differences.

\subsection{Variables}

The predictor variable $X$ is full-scale IQ as measured by age-appropriate Wechsler scales (WISC-IV for participants under 16, WAIS-IV otherwise). The outcome variable $Y$ is mean framewise displacement (FD) in millimetres, computed from the six rigid-body motion parameters estimated during fMRI preprocessing.

\subsection{Measurement Error Model}
\label{sec:sigma_model}

\paragraph{IQ measurement error ($\sigma_x$).} We derived per-observation IQ measurement error from the classical test theory formula:
\begin{equation}
    \sigma_{x,i} = \mathrm{SD}_{\mathrm{pop}} \cdot \sqrt{1 - r_{xx}},
    \label{eq:sigma_x}
\end{equation}
where $\mathrm{SD}_{\mathrm{pop}} = 15$ is the population standard deviation of IQ and $r_{xx}$ is the test-retest reliability coefficient from the relevant Wechsler technical manual. Using published reliability values from the WISC-IV \citep{Wechsler2003} and WAIS-IV \citep{Wechsler2008} manuals, this yields $\sigma_x$ values of 4.0 IQ points for children (age $< 13$), 3.4 for adolescents (age 13--15), and 3.0 for adults (age $\geq 16$). These values represent the expected standard error of measurement under standardised administration conditions.

\paragraph{FD uncertainty proxy ($\sigma_y$).} We do not observe repeated motion measurements for each subject, so we represent response-side uncertainty with a site-level proxy equal to the within-site standard deviation of mean FD across subjects:
\begin{equation}
    \sigma_{y,s} = \mathrm{SD}(\{\bar{\mathrm{FD}}_i : i \in \mathrm{site}\ s\}).
    \label{eq:sigma_y}
\end{equation}
All subjects at a given site receive the same $\sigma_{y,s}$ value. This quantity should be interpreted as a pragmatic uncertainty proxy rather than a direct measurement-error estimate, because within-site dispersion of mean FD mixes true between-subject heterogeneity with measurement noise. The advantage is that it requires no subjective tier assignment or manual grouping. Across the 19 sites, $\sigma_y$ ranged from 0.029 mm (Caltech) to 0.265 mm (NYU), with a mean of 0.129 mm.

\subsection{Probability Cloud Regression}

We introduce Probability Cloud Regression (PCR), an EM-based errors-in-variables estimator that treats each observation as a 2D Gaussian probability cloud with per-observation uncertainties $(\sigma_{x,i}, \sigma_{y,i})$. For each observation $i$:
\begin{align}
    X_i &= \xtrue_i + \varepsilon_{x,i}, \quad \varepsilon_{x,i} \sim N(0, \sigma^2_{x,i}), \\
    Y_i &= m \cdot \xtrue_i + c + \varepsilon_{y,i}, \quad \varepsilon_{y,i} \sim N(0, \sigma^2_{y,i}),
\end{align}
where $m$ and $c$ are the structural slope and intercept and $\xtrue_i$ is the latent true IQ.

\paragraph{E-step.} Given current estimates $(\mhat, \chat)$, compute the posterior mean of each latent true position:
\begin{equation}
    \xhat_i = \frac{\sigma^2_{y,i} \cdot X_i + \sigma^2_{x,i} \cdot \mhat \cdot (Y_i - \chat)}{\sigma^2_{y,i} + \mhat^2 \cdot \sigma^2_{x,i}}.
    \label{eq:estep}
\end{equation}

\paragraph{M-step.} Update regression parameters via weighted least squares with weights:
\begin{equation}
    w_i = \frac{1}{\sigma^2_{y,i} + \mhat^2 \cdot \sigma^2_{x,i}}.
    \label{eq:weights}
\end{equation}

The algorithm was initialised at the OLS solution and iterated until parameter changes fell below $10^{-8}$. Convergence occurred in 47 iterations.

When $\sigma_{x,i} = \sigma_x$ and $\sigma_{y,i} = \sigma_y$ for all $i$, PCR reduces to classical Deming regression with variance ratio $\delta = \sigma^2_y / \sigma^2_x$ \citep{Deming1943}. In the heteroscedastic case, PCR targets the same estimand as orthogonal distance regression fitted with per-observation variance matrices.

\subsection{Cross-Validation}

Leave-site-out (LOSO) cross-validation was used as the primary validation scheme. For each of the 19 sites, the model was fitted on the remaining 18 sites and used to predict mean FD for the held-out site. Overall LOSO $R^2$ was computed from concatenated predictions. This scheme tests whether a single pooled predictor of raw FD transports across sites. It is methodologically appropriate for multi-site neuroimaging because random $k$-fold cross-validation leaks site information across folds: if observations from the same site appear in both training and test sets, the model can exploit site-level structure rather than the IQ-motion relationship itself. Because held-out sites may differ in baseline motion levels as well as slope, negative LOSO $R^2$ should be interpreted as evidence against a portable pooled predictor, not by itself as proof that every within-site slope is zero.

\subsection{Sensitivity Analysis}

To assess robustness to uncertainty in the noise parameters, we evaluated PCR across an $8 \times 8$ grid of $\sigma_x$ and $\sigma_y$ multipliers. Multipliers ranged from $0.25\times$ to $3.0\times$ the empirically estimated baseline values, producing 64 configurations spanning a 12-fold range for each parameter. At each grid point, we recorded the PCR slope, the percentage by which the OLS slope magnitude exceeded the PCR slope magnitude, and the out-of-sample $R^2$.

% ============================================================
\section{Results}
\label{sec:results}
% ============================================================

\subsection{Headline Comparison}

\begin{table}[htbp]
    \centering
    \caption{OLS versus EIV-corrected estimates of the IQ-motion relationship in ABIDE-I ($n = 935$, 19 sites).}
    \label{tab:headline}
    \begin{tabular}{lcc}
        \toprule
        \textbf{Metric} & \textbf{OLS} & \textbf{PCR (EIV-corrected)} \\
        \midrule
        Slope (mm per IQ point) & $-0.00125$ & $-0.00027$ \\
        Intercept (mm) & $0.265$ & $0.118$ \\
        Bias factor$^\ast$ & $4.67\times$ overestimate & Reference \\
        LOSO $R^2$ (raw FD)$^\dagger$ & $-0.074$ & $-0.074$ \\
        \bottomrule
    \end{tabular}
    \vspace{0.4em}

    {\footnotesize $^\ast$Computed from the full-precision fitted coefficients; rounded slopes are shown for readability.}
\end{table}

The OLS slope was $-0.00125$ mm per IQ point; the PCR EIV-corrected slope was $-0.00027$ mm per IQ point, a 4.67-fold difference when computed from the full-precision coefficients underlying \Cref{tab:headline}. Both slopes are negative, indicating that higher IQ is associated with less head motion, but the magnitude of the association as estimated by OLS is nearly five times larger than the EIV-corrected estimate. Here and below, the LOSO summary should be read as the performance of a pooled predictor of raw FD rather than as a direct test that every within-site slope is null.

\subsection{Within-Site Slope Heterogeneity}

When the 19 sites were grouped by quartiles of mean FD, the within-tier OLS slopes showed a monotonic pattern (\Cref{tab:tiers}). Sites in the minimal-motion tier ($n = 290$, mean FD $= 0.074$ mm) showed essentially no IQ-motion relationship (slope $= -0.000056$). Sites in the high-motion tier ($n = 254$, mean FD $= 0.197$ mm) showed a slope 46 times steeper ($-0.00261$). The PCR-corrected slope ($-0.00027$) lies close to the within-tier slopes of the minimal and low-motion tiers, consistent with a substantially weaker pooled association in the lowest-motion sites.

\begin{table}[htbp]
    \centering
    \caption{Within-tier OLS slopes grouped by quartiles of site mean framewise displacement.}
    \label{tab:tiers}
    \begin{tabular}{lccc}
        \toprule
        \textbf{Tier} & \textbf{n} & \textbf{Mean FD (mm)} & \textbf{Within-tier slope} \\
        \midrule
        Minimal motion & 290 & 0.074 & $-0.000056$ \\
        Low motion & 163 & 0.107 & $-0.000011$ \\
        Medium motion & 228 & 0.145 & $-0.000968$ \\
        High motion & 254 & 0.197 & $-0.002605$ \\
        \bottomrule
    \end{tabular}
    \vspace{0.4em}

    {\footnotesize $^\ast$Computed from the full-precision fitted coefficients; rounded slopes are shown for readability.}
\end{table}

\begin{figure}[htbp]
    \centering
    \includegraphics[width=0.85\textwidth]{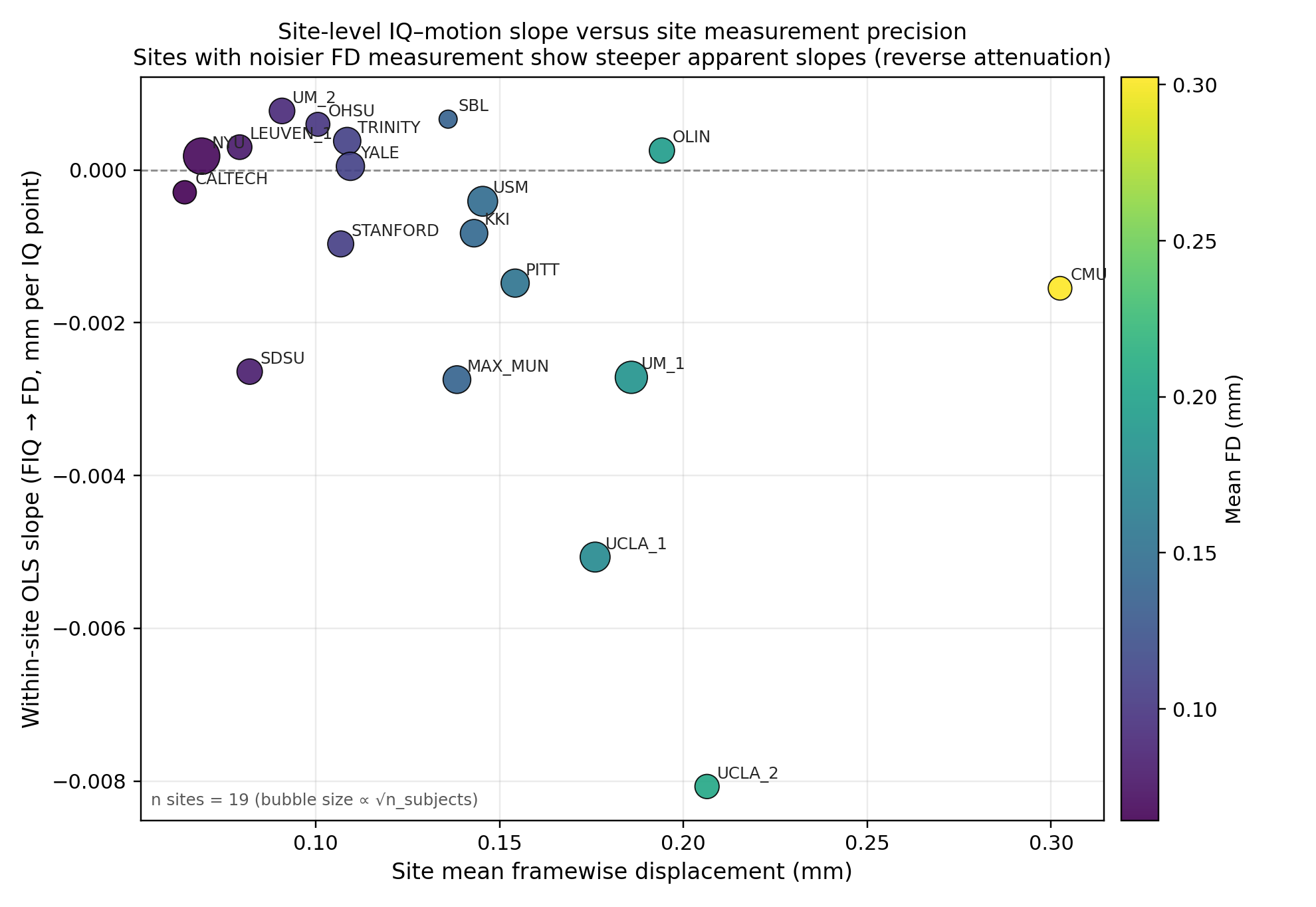}
    \caption{Site-level IQ-motion slope versus site mean framewise displacement. Each point represents one ABIDE-I site; bubble size is proportional to $\sqrt{n}$; colour encodes mean FD. Sites with larger site-level uncertainty proxies show steeper apparent slopes, consistent with subgroup-driven inflation of the pooled fit.}
    \label{fig:site_slopes}
\end{figure}

The site-level analysis (\Cref{fig:site_slopes}) confirms this pattern at the individual site level. Sites with higher mean FD (and therefore larger $\sigma_y$ proxy values) show steeper apparent IQ-motion slopes. This positive correlation between the uncertainty proxy and apparent effect size is consistent with the subgroup-driven inflation pattern studied here.

\begin{figure}[htbp]
    \centering
    \includegraphics[width=0.85\textwidth]{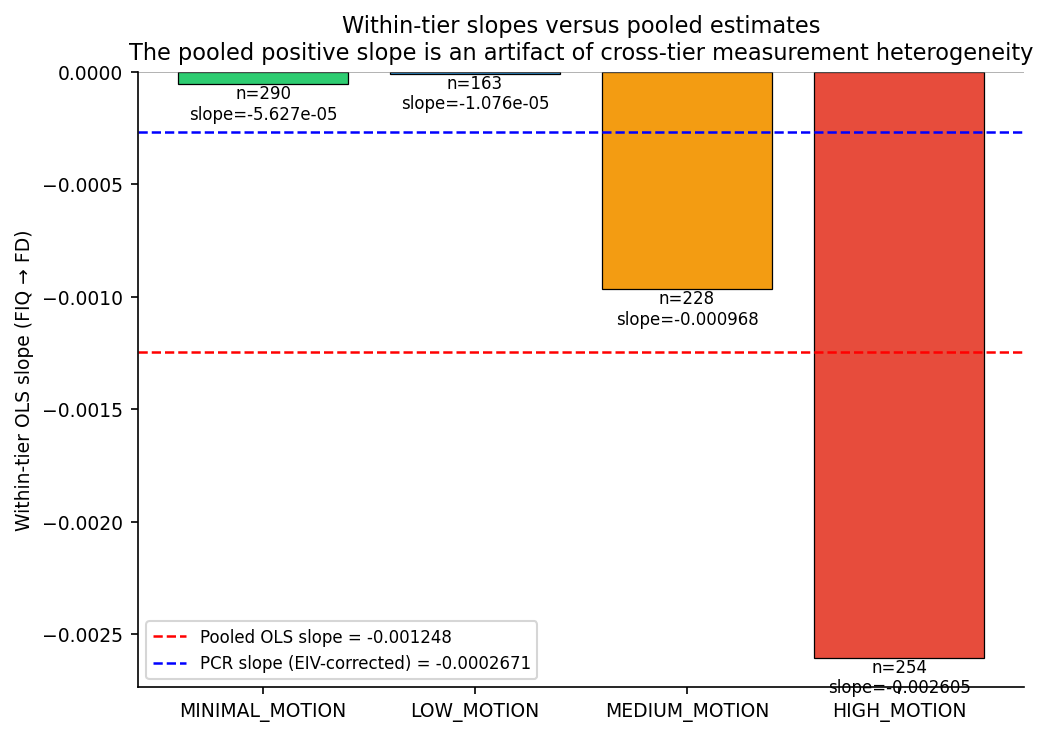}
    \caption{Within-tier OLS slopes with pooled OLS (red dashed) and PCR EIV-corrected (blue dashed) reference lines. The pooled OLS slope lies between the medium and high-motion tiers. The PCR slope lies close to the minimal and low-motion tiers, consistent with a weaker association in the lowest-motion sites.}
    \label{fig:tier_slopes}
\end{figure}

\begin{figure}[htbp]
    \centering
    \includegraphics[width=0.85\textwidth]{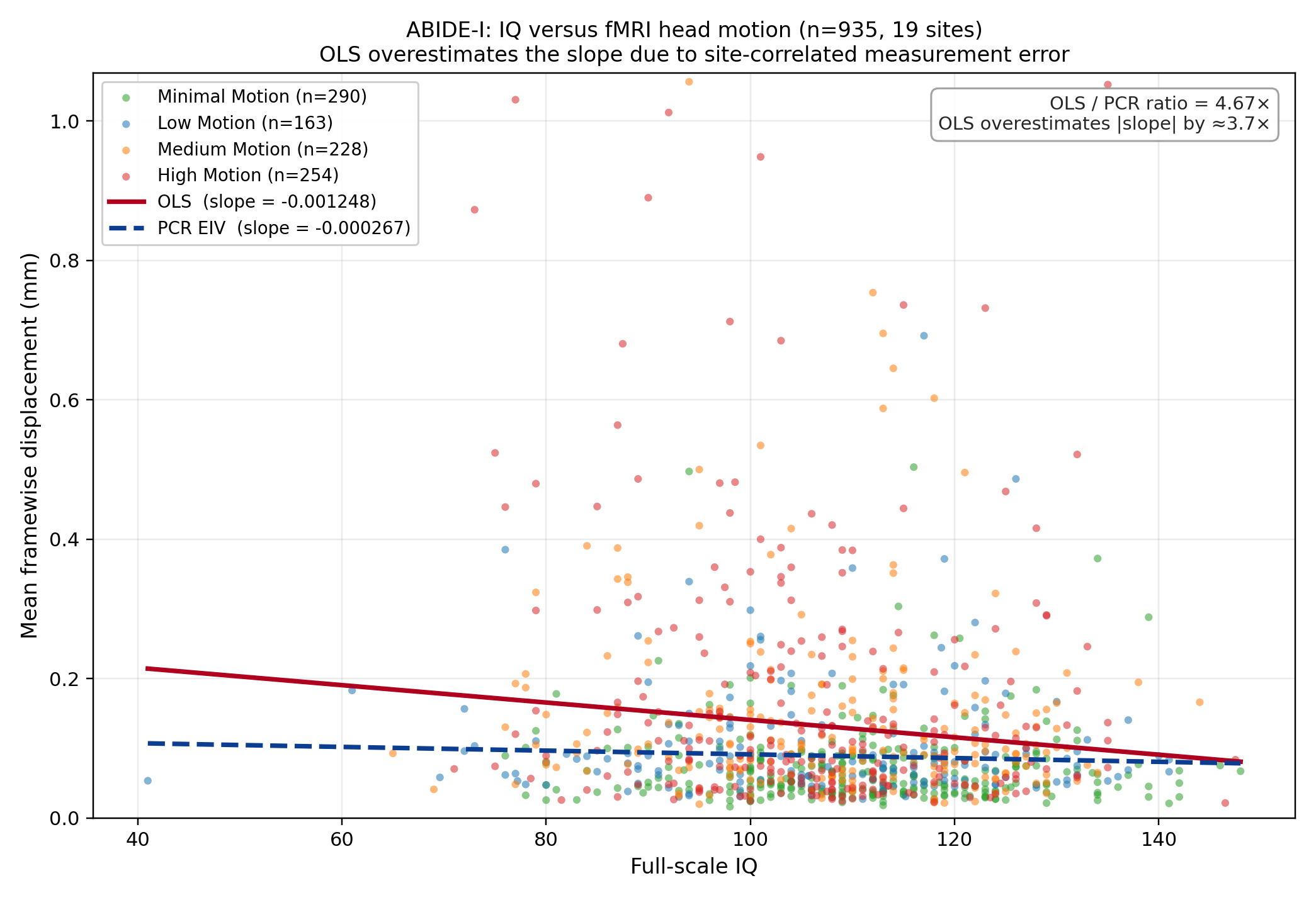}
    \caption{Full scatter of all 935 ABIDE-I subjects with OLS (solid red) and PCR EIV-corrected (dashed blue) regression lines. Points coloured by motion tier quartile.}
    \label{fig:scatter}
\end{figure}

\subsection{Leave-Site-Out Cross-Validation}

Under leave-site-out cross-validation, the pooled predictor produced negative out-of-sample $R^2$ at every single site. The overall LOSO $R^2$ was $-0.074$, and per-site values ranged from $-3.78$ (SBL, $n = 5$) to $-0.008$ (UM\_2, $n = 34$). Predicting held-out raw FD from a pooled IQ-based model trained on the other 18 sites was therefore less accurate than the naive site-level baseline.

This result shows that a single pooled predictor of raw FD does not transport cleanly across sites once site information is removed. The positive cross-validation $R^2$ that one obtains under random $k$-fold is an artifact of site information leaking between training and test folds. At the same time, LOSO here is sensitive to between-site shifts in baseline motion, so the negative $R^2$ values should not be read as direct proof that every within-site IQ-motion association is absent.

\subsection{Sensitivity Analysis}

\begin{figure}[htbp]
    \centering
    \includegraphics[width=0.85\textwidth]{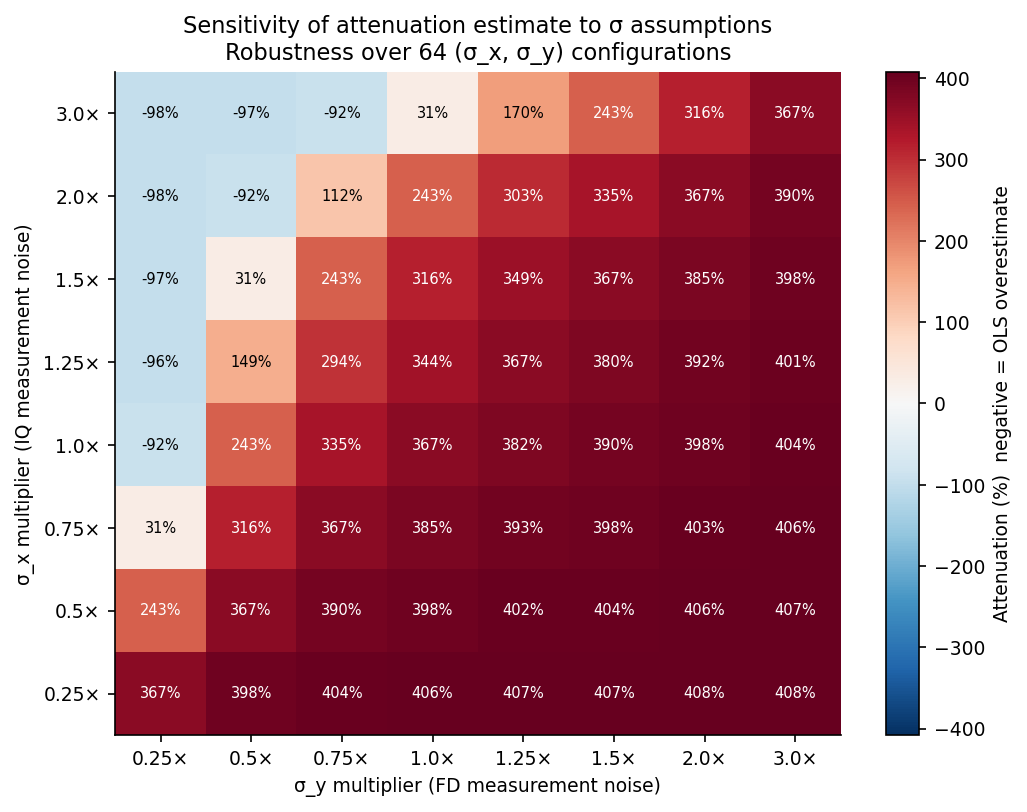}
    \caption{Sensitivity of the attenuation estimate to $\sigma_x$ and $\sigma_y$ assumptions across 64 configurations. Colour scale is diverging red-blue, centred on zero. Positive values (red) indicate OLS overestimation. The finding is robust across the full range of plausible noise assumptions.}
    \label{fig:sensitivity}
\end{figure}

Across all 64 configurations of the $8 \times 8$ sensitivity grid, the sign of the PCR-corrected slope was consistently negative (\Cref{fig:sensitivity}). The percentage by which the OLS slope magnitude exceeded the PCR slope magnitude ranged from essentially zero (at the extreme of simultaneously high $\sigma_x$ and low $\sigma_y$) to 408\% (at low $\sigma_x$ and high $\sigma_y$), but at the empirically grounded baseline ($1\times$ multipliers for both parameters) the overestimation was 367\%, corresponding to the headline 4.67$\times$ bias factor computed from the full-precision coefficients. The finding is therefore not sensitive to the precise specification of the noise parameters within the range of plausible values.

% ============================================================
\section{Discussion}
\label{sec:discussion}
% ============================================================

The central finding of this paper is that the pooled OLS estimate of the IQ-motion association in ABIDE-I can be substantially inflated when site-level response uncertainty and apparent slope magnitude co-vary. OLS estimates a slope of $-0.00125$ mm per IQ point; under the present proxy uncertainty model, the PCR estimate shrinks to $-0.00027$ mm per IQ point, a 4.67-fold reduction computed from the full-precision coefficients. The pooled predictor does not transport cleanly across sites under leave-site-out cross-validation, and the direction of the finding is robust across a wide range of noise parameter assumptions.

\subsection{The Mechanism: Reverse Attenuation}

Classical EIV theory predicts that measurement error in the predictor attenuates the OLS slope toward zero \citep{Fuller1987}. The ABIDE data produce the opposite pattern: OLS \emph{inflates} the slope away from zero. Under the present model, this happens because the response-side uncertainty proxy (captured by within-site variability in FD) is positively correlated with the magnitude of the within-site IQ-motion slope across sites. High-motion sites produce both larger uncertainty proxies and steeper apparent slopes. When these sites are pooled, the noisy, steep-slope sites dominate the OLS estimate, pulling it away from the signal that would be observed in a uniform-precision sample.

This pattern is not necessarily unique to ABIDE-I: similar subgroup-driven inflation can arise whenever subgroup structure affects both relationship magnitude and effective measurement precision. Establishing how common this regime is in other domains will require separate empirical work.

\subsection{Practical Consequences for Motion Correction}

If downstream analyses calibrate regression-based adjustments using a pooled IQ-motion slope of this kind, then the OLS estimate would imply a substantially stronger adjustment than the PCR estimate. The present study does not directly quantify connectivity-level consequences, so any claim about removed neural variance should be treated as a hypothesis rather than as a demonstrated result.

The within-tier slope analysis nonetheless provides a concrete illustration. In the two lowest-motion tiers (453 subjects, 48\% of the sample), the IQ-motion slope is essentially zero. These are the sites with the highest measurement precision, where the pooled OLS estimate is least affected by the uncertainty proxy. The entire pooled IQ-motion relationship is driven by the high-motion tiers where the proxy uncertainty is largest. This pattern motivates, but does not by itself prove, the claim that pooled OLS-based correction could be too aggressive for a substantial fraction of the sample.

\subsection{Relationship to Analytic Variability}

\citet{BotvinikNezer2020} demonstrated that researcher analytic flexibility produces substantial variability in neuroimaging results. Our finding identifies a complementary source of bias that operates \emph{before} any analytic decisions are made. Even if every team in the many-analysts exercise had used identical preprocessing and analysis pipelines, any team using OLS to estimate the IQ-motion confound would have obtained a biased estimate under the same proxy uncertainty model. This bias is not obvious from standard diagnostic checks: OLS residuals appear well-behaved, the regression is statistically significant, and random $k$-fold cross-validation yields positive $R^2$. Leave-site-out cross-validation helps reveal the portability problem by removing the site information that drives the inflated pooled estimate.

\subsection{Relationship to ComBat Harmonisation}

ComBat \citep{Johnson2007, Fortin2018} and related harmonisation methods address site effects in derived neuroimaging measures by modelling additive and multiplicative batch effects. Our analysis addresses a different problem: the bias in the regression model used to estimate relationships between phenotypic variables. ComBat could, in principle, harmonise FD values across sites before the IQ-motion regression, which would reduce between-site heterogeneity in the FD distribution and therefore in the $\sigma_y$ proxy. But ComBat does not model the measurement error in IQ, and it does not correct the OLS estimator itself. The two approaches are complementary and could be applied in sequence.

\subsection{Limitations}

First, our analysis uses phenotypic-level FD means rather than time-resolved motion traces. A stronger test would use per-subject within-session FD variability as the direct measurement noise estimate, requiring access to the raw motion parameter files rather than the summary statistics in the phenotypic CSV.

Second, the $\sigma_y$ estimate (within-site SD of mean FD) conflates between-subject variability with measurement noise. It should therefore be interpreted as a response-side uncertainty proxy rather than as a direct measurement-error estimate. If sites contain genuinely heterogeneous subject pools, the proxy overstates the measurement component; this may dampen the correction, but it also means that the present analysis is not yet a fully identified response-error model.

Third, leave-site-out cross-validation treats ``site'' as the cluster variable and tests transport of a pooled raw-FD predictor. Because held-out sites may differ in baseline motion levels as well as slope, the negative LOSO $R^2$ values should not be interpreted as direct evidence that every within-site IQ-motion slope is zero. If the real structure is better captured by scanner manufacturer, acquisition protocol, or age distribution, the LOSO result should be re-evaluated against the alternative clustering.

Fourth, the sensitivity analysis demonstrates robustness to $\sigma$ misspecification within a 12-fold range, but not to gross model violations such as non-Gaussian measurement error or within-site heteroscedasticity.

Fifth, we analysed the pooled sample without separating ASD and control groups. If the IQ-motion relationship differs between diagnostic groups, the pooled estimate may not accurately represent either group.

\subsection{Future Directions}

The most immediate next step is to replicate this analysis using the raw motion parameter files from ABIDE-I, computing per-subject within-session FD variability as a direct estimate of $\sigma_y$ rather than relying on within-site SD as a proxy. This would provide a more principled measurement error model and allow within-site heteroscedasticity to be modelled directly.

A second extension is to apply the same EIV framework to ABIDE-II and other multi-site datasets (e.g., the Human Connectome Project, UK Biobank) to test whether the reverse attenuation regime generalises beyond the ABIDE-I sample.

Third, the downstream consequences of the inflated confound estimate should be quantified directly: what happens to functional connectivity group differences when motion is regressed out using the EIV-corrected slope rather than the OLS slope? If downstream corrections based on the OLS estimate prove to be over-aggressive, re-analysing existing studies with the corrected confound estimate could recover effect sizes that were previously washed out.

% ============================================================
\section{Conclusion}
\label{sec:conclusion}
% ============================================================

In ABIDE-I, the pooled OLS estimate of the IQ-motion association is substantially larger in magnitude than the corresponding PCR estimate when site-level FD dispersion is used as a proxy for response-side uncertainty. A single pooled predictor of raw FD does not transport cleanly across sites, and the apparent association is concentrated in higher-motion sites. These results motivate measurement-uncertainty corrections in confound estimation, but the present analysis remains limited by the proxy nature of $\sigma_y$ and by the absence of a direct downstream re-analysis of connectivity measures. The broader lesson is that auxiliary regressions used to calibrate confound adjustment deserve the same scrutiny for measurement error and site heterogeneity as primary analyses.

% ============================================================
\section*{AI Disclosure}
% ============================================================

AI Disclosure. This research used multiple AI tools throughout the workflow. Claude Code (Anthropic) was used to download the ABIDE-I phenotypic dataset, implement the Probability Cloud Regression algorithm, execute all analyses, and generate figures. Claude (Anthropic) was used for literature search assistance, manuscript drafting, and iterative refinement of the experimental design, including the sensitivity analysis structure and cross-validation scheme. Prism (OpenAI) was used for manuscript review and compilation of the final LaTeX document.

The measurement error model and PCR theory was specified by the author based on expertise in statistics and psychometric measurement theory. The author independently verified all computational results, reviewed all AI-generated text for accuracy, and takes full responsibility for every claim, interpretation, and error in this manuscript.

In accordance with arXiv policy, the author confirms that generative AI tools were used as described above and that all content has been verified for accuracy. The PCR algorithm and analysis code are available at the repository listed under Data Availability for independent verification.

% ============================================================
\section*{Data Availability}
% ============================================================

The ABIDE-I phenotypic dataset is publicly available from the Preprocessed Connectomes Project (\href{http://fcon_1000.projects.nitrc.org/indi/abide/}{ABIDE project page}). Analysis code sufficient to reproduce all reported results, including the EM convergence trace, sensitivity grid, and leave-site-out predictions, is available at \texttt{https://github.com/kareem-soliman-ai/eiv-abide-replication.git} under the Creative Commons Attribution--NonCommercial--ShareAlike 4.0 International license (CC BY--NC--SA 4.0). This permits sharing and adaptation for non-commercial reproducibility and related scientific use, provided appropriate attribution is given and derivative works are distributed under the same license.

% ============================================================
\section*{Acknowledgements}
% ============================================================

The author thanks the ABIDE consortium and all contributing sites for making this dataset publicly available. The author also thanks Johan Wennemyr for independent replication of the PCR algorithm on a separate dataset.

% ============================================================
% References
% ============================================================

\bibliographystyle{plainnat}

\end{document}